\documentclass[preprintnumbers,showpacs,showkeys,preprint,secnumarabic,amssymb,amsmath,amsfonts,aps, pre]{revtex4-1}

\usepackage{amsmath}
\usepackage{amsfonts}
\usepackage{enumitem}
\usepackage{yfonts}
\usepackage{mathrsfs}

\usepackage{amssymb}
\usepackage[english]{babel}
\usepackage[pdftex]{graphicx}
\usepackage{tikz}
\usepackage{rotating}
\usepackage[pdftex,hypertexnames=true,linktocpage=true,breaklinks=true]{hyperref} 
\usepackage[hyphenbreaks]{breakurl}
\hypersetup{colorlinks=true,linkcolor=blue,anchorcolor=blue,citecolor=magenta,filecolor=blue,urlcolor=blue,bookmarksnumbered=false,pdfview=FitB}

\usepackage{esvect}

\usetikzlibrary{positioning,shapes}

\newcommand{\affA}{Aix Marseille Univ, Universit\'e de Toulon, CNRS, CPT, Marseille, France}
\newcommand{\affB}{CNRS Centre de Physique Th\'eorique UMR7332,
13288 Marseille, France}
\newcommand{\affC}{School of Science and Technology, University of Camerino, I-62032 Camerino, Italy}
\newcommand{\affD}{INFN Sezione di Perugia, I-06123 Perugia, Italy}

  \newcommand{\affG}{Quantum Biology Laboratory, Howard University, Washington, DC 20060, USA}
\newcommand{\affM}{Dipartimento di Fisica Universit\`a di Firenze, and
I.N.F.N., Sezione di Firenze, via G. Sansone 1, I-50019 Sesto Fiorentino, Italy }
\newcommand{\affP}{DSFTA, University of Siena, Via Roma 56, 53100 Siena, Italy}
\newcommand{\affQ}{Computational Biomedicine, Institute of Neuroscience and Medicine INM-9, Forschungszentrum J\"ulich, 52425 J\"ulich, Germany}

\begin{document}
\title{Electrodynamic forces driving DNA-protein interactions at large  distances}

 
\author{Elham Faraji}
\email{e.faraji@fz-juelich.de}
\affiliation{\affQ}

\author{Philip Kurian}
\email{pkurian@howard.edu}
\affiliation{\affG}

\author{Roberto Franzosi}
\email{roberto.franzosi@unisi.it}
\affiliation{\affP}
\affiliation{\affD}

\author{Stefano Mancini}
\email{stefano.mancini@unicam.it}
\affiliation{\affC}\affiliation{\affD} 
 
\author{Elena Floriani}
\email{elena.floriani@cpt.univ-mrs.fr}
\affiliation{\affA}\affiliation{\affB}

\author{Vania Calandrini}              
\email{v.calandrini@fz-juelich.de}
\affiliation{\affQ}

\author{Giulio Pettini}
\email{pettini@fi.unifi.it}
\affiliation{ \affM}

\author{Marco Pettini}
\email{marco.pettini@cpt.univ-mrs.fr}
\affiliation{\affA}\affiliation{\affB} \affiliation{\affG}

\date{\today}

\begin{abstract}
In the present paper we address the general problem of selective electrodynamic interactions between DNA and protein, which is motivated by decades of theoretical study and our very recent experimental findings (M. Lechelon et al, \textit{Sci Adv} \textbf{8,} eabl5855 (2022)). 
Inspired by the Davydov and Holstein-Fr\"{o}hlich models describing electron motion along biomolecules, and using a model Hamiltonian written in second quantization, the time-dependent variational principle (TDVP) is used to derive the dynamical
equations of the system. We demonstrate the efficacy of this {second-quantized} model for a well-documented biochemical system consisting of a restriction enzyme, \textit{Eco}RI, which binds selectively to a palindromic six-base-pair target within a DNA oligonucleotide sequence to catalyze a DNA double-strand cleavage. The time-domain Fourier spectra of the electron currents numerically computed for the DNA fragment and for the \textit{Eco}RI enzyme, respectively,
exhibit a cross-correlation spectrum with a sharp co-resonance peak. When the target DNA recognition sequence is randomized, this sharp co-resonance peak is replaced with a broad and noisy spectrum. Such a sequence-dependent charge transfer phenomenology is suggestive of a potentially rich variety of selective electrodynamic interactions influencing the coordinated activity of DNA substrates, enzymes, transcription factors, ligands, and other proteins under realistic biochemical conditions characterized by electron-phonon excitations. 
\end{abstract}
\maketitle


\section{Introduction}
Progress in molecular and cellular biology is consistently linked to a better knowledge of the structure of
and functional interplay between biomolecules such as DNA, RNA, and proteins. This structural-functional relationship is at the heart of molecular signalling which is highly organised in both time and space.
These building blocks are involved in fundamental signalling processes, highly organised in both time and space.
DNA or RNA-interacting proteins (e.g., helicases, polymerases,
nucleases, recombinases, endonucleases) modulate essential transduction processes involving nucleic acids to achieve DNA
duplication, repair, gene expression, and recombination, with such an astonishing
efficiency that raises a fundamental question from a physical point of view. With biochemical reactions mostly being
stereospecific, two (or more) reacting partners have to come in close contact and exhibit a definite spatial
orientation to initiate particular reactions. So how do the various actors in a given biochemical process
efficiently find each other? How does a protein effectively recruit the appropriate co-effector partner(s) or
selectively connect with its DNA/RNA target(s) in a crowded cyto- or nucleo-plasmic environment? In other words,
what are the physical forces that bring all these actors to the right place, in the right order, and in a reasonably
narrow window of time to sustain cellular function and ultimately cellular life? 
The classical way to tackle these issues invokes
Brownian motion or some variant, including proposals of so-called facilitated diffusion, but these alternative
explanations are largely phenomenological and lack an underlying description of the physical forces involved. At physiological temperatures, ubiquitous water molecules move chaotically and colliding with larger and heavier components produce   
a resultant force of both random intensity and direction. Hence, so Brownian reasoning goes, large
molecules move in a diffusive way throughout the cellular spaces and sooner or later shall encounter their
cognate partners. Many doubts arise when one tries to
estimate diffusion-driven activation for some of the biochemical processes mentioned above. 
In fact, free diffusion
is considerably slowed in the crowded cellular space \cite{crowd}. Moreover, the discrepancy between the observed reaction 
rates in cells and the predictions of strict random diffusion modelling have been recently questioned \cite{uno,due,tre,quattro,cinque,sei}.
{ On the other hand, for a long time it has been advocated that electrodynamic interactions acting at large distances can play an important role in bio-recognition by accelerating the encounters between cognate partners of biochemical reactions. These long-range electrodynamic interactions can be activated by different physical processes: \textit{i)} many-body dispersion forces between two thin parallel conducting cylinders \cite{ninham1,ninham2,ninham3} where an attractive force - of range much longer than the usual van der Waals $R^{-6}$ one - arises from the correlation of current fluctuations within the cylinders;  \textit{ii)}  between two neutral atoms, or small molecules, when one of the atoms is in an excited state and the transition frequencies of both atoms are similar \cite{ninham4};  \textit{iii)}
direct and inverse Hofmeister series for negatively or positively charged proteins, respectively, stemming from a complex interplay among dispersion forces, hydration, and ions in solution \cite{lonostro};
  \textit{iv)} resonant interactions between two molecules with oscillating large dipole moments entailed by collective intramolecular oscillations \cite{IJQC1}.
Long-range electrodynamic forces} could help explaining a number of phenomena in living matter,  such as the extraordinary efficiency of enzymatic reactions \cite{enzyme}, 
comprising the molecular DNA transcription machinery, certain ligand-receptor recruitments, and so on \cite{RNC,IJQC2}. 
For both technological and theoretical reasons, no formal confirmation (or refutation) of this hypothesis of electrodynamic interactions 
between biomolecules has been validated until recently. After a thorough theoretical revisitation of Fr\"ohlich's theory \cite{preto}, an  
experimental feasibility study \cite{PRE1,PRE2}, and the experimental observation of out-of-equilibrium phonon condensation in model protein in aqueous solution 
\cite{PRX} as a necessary condition \cite{preto} to activate intermolecular electrodynamic interactions, first experimental evidence
of the activation of this kind of forces has been realized \cite{SciAdv}.  {A recent molecular dynamics investigation with optically excited chromophores in a model protein in thermal equilibrium with the aqueous environment has demonstrated the complex interplay of chromophore, protein, and solvent degrees of freedom in producing the observed terahertz modes~\cite{kazizi}.}
\smallskip


Within this newly opened field, the aim of the present work is
to adapt and combine an approach inspired by the Resonant Recognition Model (RRM) \cite{veljkovic1985,Cosic0} and a theoretical treatment of intermolecular interactions mediated by dipolar waves in the aqueous environment \cite{kurian2018}.
This is in line with the attempt to understand whether intermolecular electrodynamic
interactions are implicated under different conditions of activation, paying particular attention 
to resonance effects, which are crucial for selective recruitment of the cognate partners
of a biochemical reaction.
Then, we show that  the time-domain Fourier spectra of the electron currents numerically computed for the DNA fragment and for the EcoRI enzyme, respectively, exhibit a cross-correlation spectrum with a sharp co-resonance peak. Instead, when the target
DNA recognition sequence is randomized, this sharp co-resonance peak is replaced with a broad and noisy spectrum.

This is in line with the attempt to understand 
 whether intermolecular electrodynamic interactions {are implicated} under different conditions of activation, {paying particular attention to} resonance {effects, which are crucial} for selective recruitment of the cognate partners of a biochemical reaction. 

 This paper is organized as follows.  In Section II we define the model used to describe the electron motions along the DNA fragment and the EcoRI enzyme, respectively. Section III contains the definition of the physical parameters used in the numerical simulations of the model equations. The results of these numerical simulations are then reported in Section IV. The possibility of activating water-mediated DNA-\textit{Eco}RI interaction {through many-body dispersion and field theory approaches} is discussed in Section V. Finally, in Section VI some concluding remarks are made.
\section{Definition of a dynamical model and its solution}

In our recent work \cite{scirep} we found a rich phenomenology of the current flowing along a DNA fragment under the action of an external source of energy: according to the excitation site and energy the resulting electron current can display either a broad frequency spectrum or a sharply peaked frequency spectrum. This suggested  to tackle the DNA-enzyme interaction by borrowing the Resonant Recognition Model (RRM) philosophy  with the aid of an explicit modelling of the electronic motions along the backbones of interacting DNA-protein biomolecules. In order to describe these electronic motions and their electrodynamic interactions we resort to a model partly borrowed from the standard Davydov and Holstein-Fr\"ohlich models that have been originally introduced to account for electron-phonon interaction \cite{standard,froehpolaron,holstein}. Thus, to separately model the electrons moving along a given DNA sequence and along the backbone of a DNA-interacting enzyme (we will consider the EcoRI restriction enzyme),  the following common Hamiltonian operator for both EcoRI enzyme and DNA is assumed 
   \begin{eqnarray}\label{H}
   \hat{H} =\hat{H}_{el}+\hat{H}_{ph}+\hat{H}_{int},
   \end{eqnarray}
with
   \begin{equation}\label{Hex}
   \hat{H}_{el}=\sum_{n=1}^{N}
   \Big[ E_{0}\hat{B}_{n}^{\dag} \hat{B}_{n}+\epsilon \langle\hat{B}_{n}^{\dag } \hat{B}_{n}\rangle \hat{B}_{n}^{\dag } \hat{B}_{n}+J_n(\hat{B}_{n}^{\dag} \hat{B}_{n+1}+\hat{B}_{n}^{\dag} \hat{B}_{n-1})\Big],
      \end{equation}
  \begin{equation}\label{Hph}
       \hat{H}_{ph}=\frac{1}{2}\sum_{n}\Big[\frac{\hat{p}_{n}^2}{M_n}+\Omega_{n}(\hat{u}_{n+1}-\hat{u}_{n})^2+\frac{1}{2}\mu (\hat{u}_{n+1}-\hat{u}_{n})^{4}\Big],
         \end{equation} 
   \begin{equation}\label{Hin}
          \hat{H}_{int}=\sum_{n}\chi_{n}(\hat{u}_{n+1}-\hat{u}_{n})\hat{B}_{n}^{\dag}\hat{B}_{n}.
            \end{equation} 
in which $ \hat{H}_{el}$ and  $\hat{H}_{ph}$ are respectively the electronic and phononic Hamiltonians and $\hat{H}_{int}$ indicates the electron-phonon interaction term. At variance with the models investigated in Refs.\cite{scirep,Faraji}, the coupling parameters $J_n$ and $\chi_n$ are assumed to be site-dependent.  
Considering only a longitudinal chain of amino acids (or nucleotides), $\hat{B}_n$ and $\hat{B}^\dag_n$ denote the lowering and raising operators between the lattice site $n\in \{1,2,\dots ,N\}$ labelling the amino acids along the EcoRI enzyme (or nucleotides along a DNA).
The parameter $E_0$ defines the initial excitation energy of the electron according to the initial form of the electronic state vector.
The nonlinear constant $\epsilon$ is the coupling energy of the interaction between the moving electron along the chain with the electrons of the substrate of amino acids (or nucleotides). The coupling parameter $J_n$ is a site-dependent tunnelling term of electrons across two nearest neighbouring amino acids (or nucleotides).

The momentum and position operators $\hat{p}_{n}$ and $\hat{u}_{n}$ of the vibronic Hamiltonian determine the longitudinal displacements of the $n$-th phonon in the sequence of amino acids (or nucleotides) from their equilibrium position and the coupling term $\Omega_n$ denotes the site-dependent spring parameter of two neighbouring sites. $M_n$ is the mass of the $n$-th amino acid of EcoRI enzyme sequence ( or nucleotide of a DNA segment) and the nonlinear coupling constant $\mu$ implies phonon-phonon interaction, absent in the harmonic approximation. Finally, the parameter $\chi_{n}$ of the interaction Hamiltonian is the $n$-th site-dependent electron-phonon coupling. \\
The wave function $|\psi(t)\rangle$ at any time $t$ may be written in the Davydov ansatz by the following factorization
  \begin{eqnarray}\label{psi}
|\psi(t)\rangle=|\Psi(t)\rangle|\Phi(t)\rangle,
   \end{eqnarray} 
with the normalization condition $\langle\psi(t)|\psi(t)\rangle=1$. The state vector $|\Psi(t)\rangle$ describes a single quantum excitation of an electron propagating along a protein chain of N amino acids ( or a DNA sequence of $N$ nucleotides)
  \begin{eqnarray}\label{Psi}
|\Psi(t)\rangle=\sum_{n}C_{n}(t)\hat{B}_{n}^{\dag}|0\rangle_{el},
   \end{eqnarray}
 in which $|0\rangle_e$ is the electronic vacuum state, and $ |\Phi(t)\rangle$ is the vibronic wave function
  \begin{eqnarray}\label{Phi}
 |\Phi(t)\rangle=e^{-i/ \hbar\sum[\beta_{n}(t)\hat{p}_{n}-\pi_{n}(t)\hat{u}_{n}]}|0\rangle_{ph},
   \end{eqnarray}
 for which the expectation values for longitudinal displacement $\hat{u}_{n}$ and momentum $\hat{p}_{n}$ are, respectively, $\langle\Phi|\hat{u}_{n}|\Phi\rangle =\beta_{n}(t)$ and $\langle\Phi|\hat{p}_{n}|\Phi\rangle =\pi_{n}(t)$.
According to the time-dependent variation principle (TDVP), we define a phase factor $(S(t)\in \mathbb{R})$ and set a new wave function $|\phi(t)\rangle$ from Eq.(\ref{psi}) as $|\phi(t)\rangle=e^{iS(t)/\hbar}|\psi(t)\rangle$ satisfying the normalization $\langle \phi(t)|\phi(t)\rangle=1$. Integrating the quantum Schr\"{o}dinger equation, $i\hbar\langle\phi(t) | \partial_t|\phi(t)\rangle=\langle\phi(t) |\hat{H}|\phi(t)\rangle$ leads to $S(t)= \int_{0}^{t} L(t') dt'$ which can be supposed to be the classical Lagrangian associated to the system
\begin{eqnarray}\label{LL}
L(t)=i\hbar \langle\psi(t) |\partial_t|\psi(t)\rangle-\langle\psi(t) |\hat{H}|\psi(t)\rangle.
   \end{eqnarray} 
Now, TDVP which is equivalent to the least action principle reads 
           \begin{eqnarray}\label{stationary}
\delta S(t)=\delta \int_{0}^{t} L(t')dt'=0.
   \end{eqnarray} 
Then, from the wave function (\ref{psi}) and Lagrangian (\ref{LL}) we have
    \begin{eqnarray}
L=\sum_{n}\left\{i\hbar \dot{C}_{n}(t)C_{n}^{*}(t)+{1\over2}\Big(\pi_{n}(t)\dot{\beta}_{n}(t)-\dot{\pi}_{n}(t)\beta_{n}(t)\Big)-H(C_{n},C_{n}^*,\beta_{n},\pi_{n})\right\},
   \end{eqnarray}
   in which $H(C_{n},C_{n}^*,\beta_{n},\pi_{n})= \langle\psi(t) |\hat{H}|\psi(t)\rangle$. Hence and with the stationary action \eqref{stationary} one  obtains
\begin{eqnarray}
\delta S(t)&=\sum_{n}\Big\{ i\hbar\Big(-\dot{C}_{n}^{*}(t)\delta C_{n}(t)+\dot{C}_{n}(t)\delta C_{n}^{*}(t)\Big)+\dot{\beta}_{n}(t)\delta \pi_{n}(t)-\dot{\pi}_{n}(t)\delta \beta_{n}(t)\nonumber\\
&-(\partial_{C_{n}}H)\delta C_{n}-(\partial_{C_{n}^{*}}H)\delta C_{n}^{*}-(\partial _{\beta_{n}}H)\delta \beta_{n}-(\partial_{\pi_{n}}H)\delta \pi_{n}\Big\}=0,
   \end{eqnarray}
 which gives the equations 
           \begin{eqnarray}\label{hc}
i\hbar\dot{C}_{n}&=&\partial_{C^*_{n}} H\nonumber\\
\dot{\beta}_{n}&=&\partial _{\pi_{n}} H\nonumber\\
\dot{\pi}_{n}&=&-\partial _{\beta_{n}} H.
   \end{eqnarray}
Using the expectation value of the Hamiltonian 
      \begin{eqnarray}\label{Hnew}
\langle \psi|\hat{H}|\psi \rangle&=\sum_{n}\Big[E_{0}|C_{n}|^{2}+\epsilon|C_{n}|^{4}+ J_n(C_{n}^{*}C_{n+1}+C_{n+1}^{*}C_{n})\nonumber\\
&+ \frac{1}{ 2}\Big(\frac{1}{M_n}\pi_{n}^2+\Omega_{n} (\beta_{n+1}-\beta_{n})^2+{1\over 2}\mu (\beta_{n+1}-\beta_{n})^4\Big)\nonumber\\
&+\chi_n(\beta_{n+1}-\beta_{n})|C_{n}|^2\Big],
   \end{eqnarray}
and Eqs. (\ref{hc}) the equations of the motion are found to be 
 \begin{eqnarray}\label{dynamicaleqs}
  i\hbar\dot{C}_{n}&=&\Big(E_{0}+2\epsilon|C_{n}|^{2}+\chi_n(\beta_{n+1}-\beta_{n})\Big)C_{n}
 +J_{n} C_{n+1}+J_{n-1}C_{n-1},\nonumber\\
 M_{n}\ddot{\beta}_{n}&=&\Omega_{n}\beta_{n+1}+\Omega_{n-1}\beta_{n-1}-\Omega_{n-1}\beta_{n}-\Omega_{n}\beta_{n}+\chi_n |C_{n}|^2-\chi_{n-1} |C_{n-1}|^2\nonumber\\
&+&\mu \Big((\beta_{n+1}-\beta_{n})^{3}-(\beta_{n}-\beta_{n-1})^{3}\Big).
    \end{eqnarray}
    It is worth noting that the dynamical equations worked out by means of the TDVP are formally classical but give the time
    evolution of actual quantum expectation values.
    
  \section{Physical parameters for the numerical computations}\label{numericalcomputations}
  
  We need to determine the physical values of the coupling parameters of the Hamiltonian to perform our numerical simulations. To this aim we borrow  from Ref.\cite{Cosic,pseudopot} the  interaction energy of an electron with any given amino acid as per table I, and  the  interaction energy of an electron with any given nucleotide as per table II.   
\begin{figure}[h!]
\includegraphics[width=0.85\columnwidth]{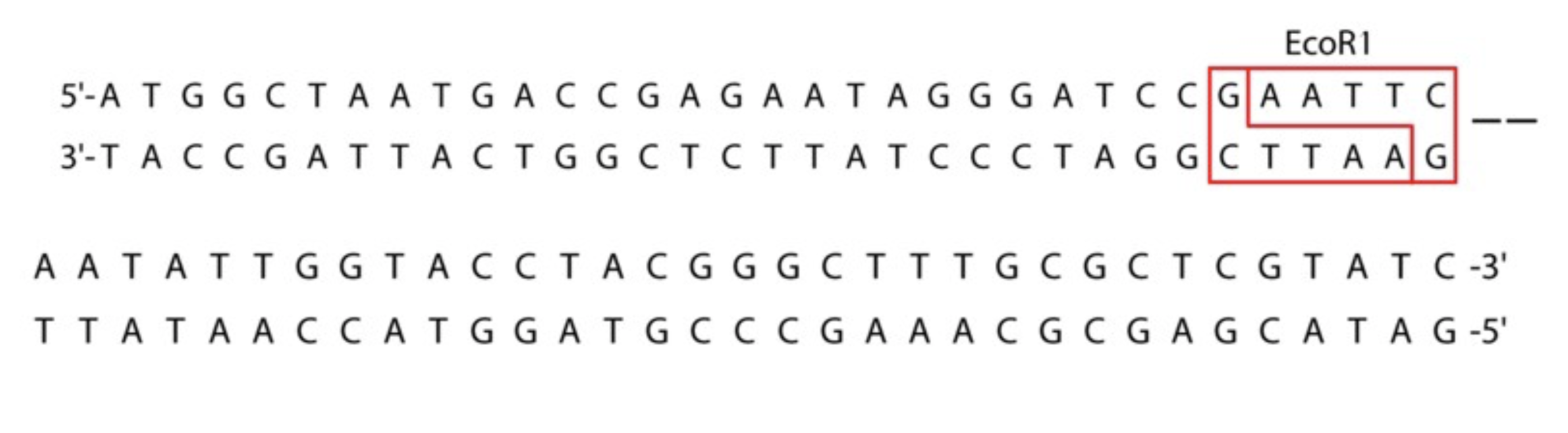}
\includegraphics[width=0.85\columnwidth]{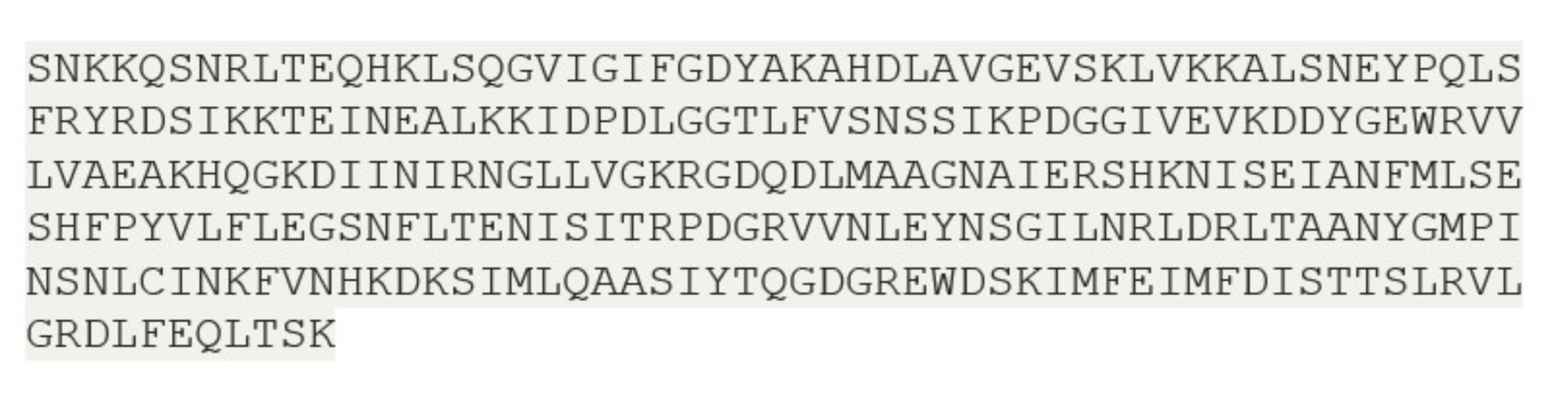}
   \caption{Top: DNA sequence of an oligonucleotide (66 base pairs) containing a  target cleavage subsequence of the \textit{Eco}RI enzyme. Bottom: Amino acid sequence of the \textit{Eco}RI enzyme (1QC9 of PDB).} 
  \label{EcoRI seq1}
 \end{figure}
 
  \begin{center}
\begin{table}[htbp]\label{energies}
\centering
\begin{tabular}{|c|c|c||c|c|c|}
\hline
          \textbf{Nucleotide} &\textbf{EIIP Ry}&\textbf{EIIP eV}& \textbf{Nucleotide} &\textbf{EIIP Ry}&\textbf{EIIP eV}
        \\  
        \hline
       A&0.1260& 1.7143&     T&0.1335&  1.8164\\
     G&0.0806&  1.0966&      C&0.1340&  1.8232   \\    
   \hline              
\end{tabular}
\caption{Electron-Ion interaction potential (EIIP) values for nucleotides adenine (A), thymine (T), guanine (G), and cytosine (C). From Ref.\cite{Cosic}.}
\end{table}
\end{center}
\vspace{ -1.5cm}
  \begin{center}
\begin{table}[htbp]
\centering

\label{electron-amino-acids}
\begin{tabular}{|c|c|c||c|c|c|}
\hline
          \textbf{Amino acid} &\textbf{EIIP Ry}&\textbf{EIIP eV}& \textbf{Amino acid} &\textbf{EIIP Ry}&\textbf{EIIP eV}
        \\  
        \hline
      Leu&0.0000& 0.0000&     Tyr&0.0516&  0.7017\\
      Ile  &0.0000&  0.0000&   Trp&0.0548&  0.7452\\
      Asn &0.0036& 0.0489&    Gln&0.0761&  1.0349\\
      Gly &0.0050& 0.0680&      Met&0.0823&  1.1192\\
      Val &0.0057& 0.0775&      Ser&0.0829& 1.1274\\
      Glu&0.0058&  0.0788&    Cys&0.0829&  1.1274\\
        Pro&0.0198&  0.2692& Thr&0.0941&  1.2797\\
       His&0.0242&  0.3291&   Phe&0.0946&  1.2865\\
       Lys&0.0371&  0.5045&  Arg&0.0959&  1.3042\\
        Ala&0.0373& 0.5072&   Asp&0.1263& 1.7176\\
   \hline              
\end{tabular}
\caption{Electron-Ion interaction potential (EIIP) value for 20 amino acids. From Ref.\cite{Cosic}.}
\end{table}
\end{center}
The electron moving with the initial energy $E_0$ experiences a periodic sequence of square potential barriers of different heights and of the same width $a$ - the average distance between two nearest neighboring sites - by tunneling across the chain of amino acids constituting a protein or the sequence of nucleotides composing DNA. The value of distance $a$ is $4.5 \mathring{A}$ in EcoRI  enzyme and $3.4 \mathring{A}$ in DNA fragment. We can then estimate roughly the electron tunneling term as $J_{n}=E_{0}T_{n,n+1}$, by introducing the transmission coefficient $T_{n,n+1}$ from the probability $P(n\rightarrow n\pm 1)$ of tunneling from one potential barrier to the nearest one.
This is done  as follows:
  \begin{itemize}
 \item Case 1: $E_{0}<E_{n+1}$ 
   \begin{equation}\label{barrier}
   T_{n,n+1}= \left[ 1 + \frac{ E^{2}_{n+1} \sinh^2(\beta_{n+ 1} a)}{4 E_0(E_{n+ 1}  - E_0)}\right]^{-1},
   \end{equation}
  where $\beta_{n+1} =[2 m_e(E_{n+ 1} - E_0)/\hbar^2]^{1/2}$.
  \item  Case 2:  $E_{0}>E_{n+1}$
      \begin{equation}\label{barrier2}
 T_{n,n+1}= \left[ 1 + \frac{ E^{2}_{n+1} \sin^2(\beta_{n+ 1} a)}{4 E_0( E_0-E_{n+1})}\right]^{-1},
      \end{equation}
in which $\beta_{n+1} =[2 m_e(E_0-E_{n+1})/\hbar^2]^{1/2}$. 
  \end{itemize}
 Here $m_e$ is the mass of electron and $E_{n+1}$ are the potential interaction energies between the electrons in motion and the local amino acids ( or nucleotides).
Moreover, in a rough estimation we set $\chi_{n}=d E/d x =(E_{n+1}-E_{n})/a$ as the site-dependent electron-phonon coupling.\\
In order to perform the numerical simulations, the dimensionless expectation value of the Hamitonian (\ref{Hnew}) and of the dimensionless equations of motion (\ref{dynamicaleqs}) are found by rescaling time $t=\omega^{-1}\tau$ and length $\beta_{n}=Lb_n$ where $L=(\hbar \omega^{-1}M_{n}^{-1})^{1/2}$. We then obtain 
         \begin{align}\label{Hsimpli}
\langle \psi|\hat{H}|\psi \rangle&=\sum_{n}\Big[E'|C_{n}|^{2}+\epsilon'|C_{n}|^{4}+ J_{n}'(C_{n}^{*}C_{n+1}+C_{n+1}^{*}C_{n})\nonumber\\
&+ {1\over 2}\Big(\dot{b}^{2}_{n}+ \Omega_{n}'(b_{n+1}-b_{n})^2+{1\over 2}\mu'(b_{n+1}-b_{n})^4\Big)\nonumber\\
&+\chi_{n}'(b_{n+1}-b_{n})|C_{n}|^2\Big],
   \end{align}
and 
\begin{eqnarray}\label{dimdynamics}
  i\frac{d{C}_{n}}{d\tau}&=&\Big(E'+2\epsilon'|C_{n}|^{2}+\chi_{n}'(b_{n+1}-b_{n})\Big)C_{n}
 +J_{n}'C_{n+1}+J_{n-1}'C_{n-1},\nonumber\\
\frac{d^{2}{b}_{n}}{d\tau^2}&=&\Omega_{n}'b_{n+1}+\Omega_{n-1}' b_{n-1}-\Omega_{n-1}'b_{n}-\Omega_{n}'b_{n}+\chi_{n}'|C_{n}|^2-\chi_{n-1}'|C_{n-1}|^2\nonumber\\
&+&\mu'\Big[(b_{n+1}-b_{n})^3-(b_{n}-b_{n-1})^3\Big],
   \end{eqnarray}
where the dimensionless parameters are
   \begin{eqnarray}\label{dimparameter}
   E'&=&{E_{0} \over \hbar \omega};\;\;\;\;\;\;\; \epsilon'={\epsilon \over \hbar \omega};\;\;\;\;\;\;\; J_{n}'={J_{n}\over \hbar \omega};\;\;\;\;\;\;\; \nonumber\\
    \chi_{n}'&=&{\chi_{n} \over \sqrt{\hbar M_{n}\omega^{3}}};\;\;\;\;\; \Omega_{n}'={\Omega_{n}\over M_{n} \omega^2};\;\;\;\;\;\;\mu'={\mu \hbar \over M_{n}^{2}\omega^{3}}\ .
   \end{eqnarray}

   The sound speed of amino acids is $V\sim 4$ Km/s from \cite{standard,Cruzeiro}, and the one of nucleotides is $V=1.69$ Km/s from \cite{Hakim} (neglecting small local variations due to the different masses of the amino acids or the nucleotides). We apply two different analyses for computing the spring parameter in our simulations. First, we consider the known speed of sound $V=a(\Omega_{n}/M_{n})^{1/2}$ leading to the constant dimensionless parameter $\Omega'=V^{2}/a^{2}\omega^2$ from (\ref{dimparameter}) where $\Omega'=0.79$ for amino acids and $\Omega'=0.25$ for nucleotides. Second,  from \cite{standard} we borrow the spring constant of amino acids as $\Omega=18.3 N/m$ whence, after Eq. (\ref{dimparameter}), the site-dependent dimensionless quantities $\Omega'_{n}=1.83/m_{n}$ are obtained. Third, we assume the average spring constant $\Omega=V^{2}\langle M\rangle/a^2$ of DNA - in which $\langle M\rangle$ is the average masse of the nucleotides - whence the dimensionless site-dependent parameters $\Omega_{n}'=0.48/m_{n}$ for nucleotides follow. The expression $m_{n}$ represents the dimensionless mass of amino acids and nucleotides.\\
    In order to perform numerical integration of the dynamical equations it is useful to introduce the variables
           \begin{eqnarray}
q_{n}={C_{n}+C_{n}^{*}\over \sqrt{2}}, \qquad p_{n}={C_{n}-C_{n}^{*}\over i\sqrt{2}},
   \end{eqnarray}
which allows to rewrite Eqs.(\ref{dimdynamics}) as
\begin{align}
           \dot{q}_{n}&=\Big[E'+{\epsilon' \over 2}(q_{n}^{2}+p_{n}^{2})+\chi'(b_{n+1}-b_{n})\Big]p_{n}+J_{n}'p_{n+1}+J_{n-1}'p_{n-1},\label{differential28}\\
\dot{p}_{n}&=-\Big[E'+{\epsilon' \over 2}(q_{n}^{2}+p_{n}^{2})+\chi'(b_{n+1}-b_{n})\Big]q_{n}+J_{n}'q_{n+1}+J_{n-1}'q_{n-1}\Big], \label{differential29}\\
\ddot{b}_n&=\Omega'(b_{n+1}+b_{n-1}-2b_{n})+ {1\over 2}  \Big(
\chi_{n}'(q_{n}^{2}+p_{n}^{2})-\chi_{n-1}'(q_{n-1}^{2}+p_{n-1}^{2})\Big)\nonumber\\
&+\mu'\Big[(b_{n+1}-b_{n})^3-(b_{n}-b_{n-1})^3\Big]\label{differentialqp}.
   \end{align}
Denoting the r.h.s of the above equation (\ref{differentialqp}) by $\mathcal{B}_n[\textbf{b}(t), \textbf{q}(t), \textbf{p}(t)]$ and writing the second time derivatives of $b_n$ in finite differences, equation (\ref{differentialqp}) reads as $b_n(t+\Delta t) = 2 b_n(t) - b_n(t - \Delta t) + (\Delta t)^2 \mathcal{B}_n[\textbf{b}(t), \textbf{q}(t), \textbf{p}(t)]$; therefore 
\begin{eqnarray}\label{bienne}
\dot{b}_n&=& \pi_n\ , \nonumber\\
\dot{\pi}_n&=& \mathcal{B}_n[\textbf{b}(t), \textbf{q}(t), \textbf{p}(t)]\ .
\end{eqnarray}
Furthermore, we denote by $\mathcal{Q}_n[\textbf{b}(t), \textbf{q}(t), \textbf{p}(t)]$ and $\mathcal{P}_n[\textbf{b}(t), \textbf{q}(t), \textbf{p}(t)$, respectively,  the r.h.s. of Eqs.\eqref{differential28} and \eqref{differential29}, and perform the numerical integrations by combining a finite differences scheme and a leap-frog scheme as follows
   \begin{eqnarray}\label{eqdinamiche}
q_{n}(t+\Delta t)&=& q_{n}(t)+\Delta t\ \mathcal{Q}_n[\textbf{b}(t), \textbf{q}(t), \textbf{p}(t)],\nonumber\\
p_{n}(t+\Delta t)&=& p_{n}(t)+\Delta t\ \mathcal{P}_n[\textbf{b}(t), \textbf{q}(t), \textbf{p}(t),\nonumber\\
b_{n}(t+\Delta t)&=& b_{n}(t)+\Delta t\ \pi_n(t),\nonumber\\
\pi_{n}(t+\Delta t)&=& \pi_{n}(t)+\Delta t\ \mathcal{B}_n[\textbf{b}(t+\Delta t), \textbf{q}(t+\Delta t), \textbf{p}(t+\Delta t)] .
   \end{eqnarray}
   The integration scheme for $b_n(t)$ and $p_n(t)$ is a symplectic one, meaning that all the Poincar\'e invariants of the associated Hamiltonian flow are conserved, among whom there is energy. We can not apply the simple leap-frog scheme to the equations for $\dot{q}_n(t)$ and $\dot{p}_n(t)$, since the r.h.s. of the equations for $\dot{q}_n(t)$ explicitly depend on $q_n(t)$ and $b_n(t)$; therefore, we integrate the first two equations in (\ref{eqdinamiche}) with an Euler predictor-corrector to get
 \begin{eqnarray}\label{eqdinamicheBis}
   q^{(0)}_{n}(t+\Delta t)&=& q_{n}(t)+\Delta t\ \mathcal{Q}_n[\textbf{b}(t), \textbf{q}(t), \textbf{p}(t)],\nonumber\\
   p^{(0)}_{n}(t+\Delta t)&=& p_{n}(t)+\Delta t\ \mathcal{P}_n[\textbf{b}(t), \textbf{q}(t), \textbf{p}(t)],\nonumber\\
   q^{(1)}_{n}(t+\Delta t)&=& q_{n}(t)+\frac{\Delta t}{2}\left\{ \mathcal{Q}_n[\textbf{b}(t), \textbf{q}(t), \textbf{p}(t)] +  \mathcal{Q}_n[\textbf{b}(t), \textbf{q}^{(0)}(t+\Delta t), \textbf{p}^{(0)}(t+\Delta t)]\right\},\nonumber\\
   p^{(1)}_{n}(t+\Delta t)&=& p_{n}(t)+\frac{\Delta t}{2}\left\{\mathcal{P}_n[\textbf{b}(t), \textbf{q}(t), \textbf{p}(t)] +  \mathcal{P}_n[\textbf{b}(t), \textbf{q}^{(0)}(t+\Delta t)), \textbf{p}^{(0)}(t+\Delta t)]  \right\},\nonumber\\
   b_{n}(t+\Delta t)&=& b_{n}(t)+\Delta t\ \pi_n(t),\nonumber\\
   \pi_{n}(t+\Delta t)&=& \pi_{n}(t)+\Delta t\ \mathcal{B}_n[\textbf{b}(t+\Delta t), \textbf{q}^{(1)}(t+\Delta t), \textbf{p}^{(1)}(t+\Delta t)].
  \end{eqnarray}
The integration of half of the set of the dynamical equations \eqref{eqdinamiche} by means of a symplectic algorithm, and half of the equations by means of the Euler predictor-corrector \eqref{eqdinamicheBis} results in a very good conservation of total energy without any shift- just with zero-mean fluctuations around a given value fixed by the initial conditions - by considering sufficiently small integration time steps $\Delta t$. We also need to define the initial states of electron and phonon independently of the specific physical excitation mechanism. The electron wavefunction (\ref{Psi}) is described by the amplitudes $C_{n}(t=0)$ centered at the excitation site $n=n_{0} $ and distributed at time $t=0$ \cite{standard} as 
 \begin{eqnarray}\label{exc}
C_{n}(t=0)&=& {1\over \sqrt{8 \sigma_{0}}}{\rm sech}\Big({n-n_{0}\over 4\sigma_{0} }\Big)
    \end{eqnarray}\\
where $\sigma_{0}$ specifies the amplitude width. Concerning the phonon part of the system, we consider a thermalized macromolecule EcoRI enzyme and DNA fragment at room temperature $T =310^{\circ} K$. At thermal equilibrium, average kinetic and potential energies per degree of freedom are equal, and the total energy is equally shared among all the phonon modes. Accordingly, the displacements and the associated velocities have been initialized with random values of zero-mean at $t=0$, then in a dimensionless form we have
     \begin{equation}\label{thermequil}
  \langle\vert b_{n}(0)\vert\rangle_n=\sqrt{\frac{k_{B}T}{\hbar \omega \Omega'}} ;\;\;\;\;\;\;\;\;\;\;\;\;\;\;\;\;\; \langle\vert \pi_{n}(0)\vert\rangle_n=\sqrt{\frac{k_{B}T}{\hbar \omega}}.
      \end{equation}
Periodic boundary conditions have been used for both the  electron and phonon part of the DNA-EcoRI interacting system and the frequency has been assumed to be $\omega=10^{13}$s$^{-1}$.

\section{Numerical Results}
\begin{figure}[h!]
\includegraphics[width=0.45\columnwidth]{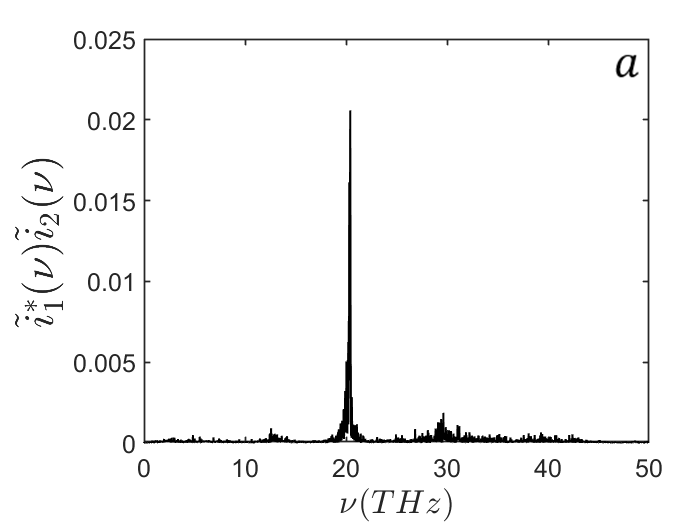}
\includegraphics[width=0.45\columnwidth]{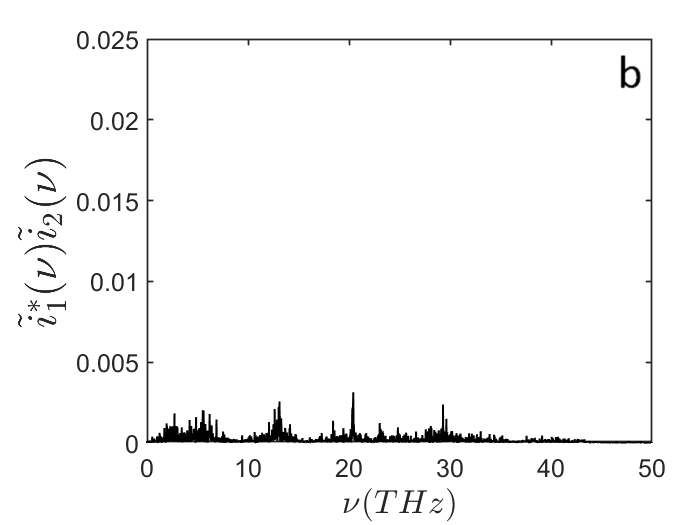}
\includegraphics[width=0.45\columnwidth]{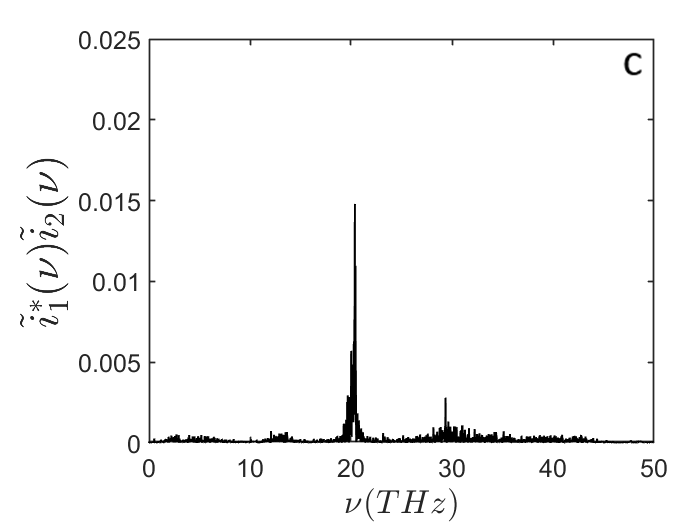}
\includegraphics[width=0.45\columnwidth]{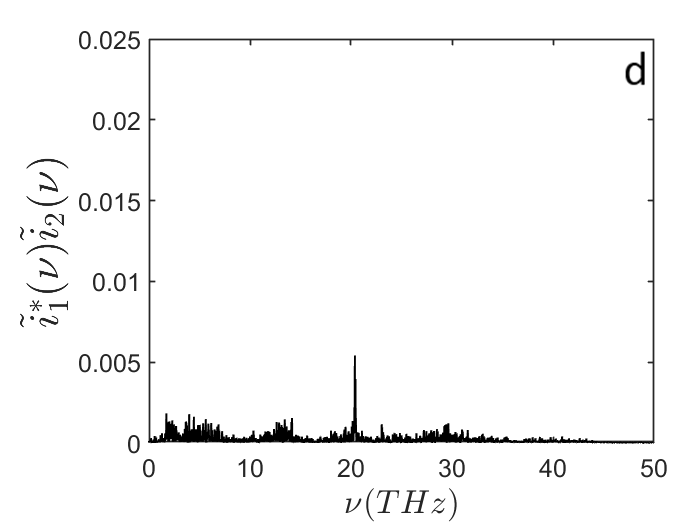} 
   \caption{The cross frequency spectrum of the interaction between DNA strand with $N_{1}=66$ nucleotides and the EcoRI enzyme with $N_{2}=276$ amino acids for the initial conditions $T=310^{\circ}K$,  $N_{0,1}=N/2$,  $N_{0,2}=N/3$, $E'_{1,0}=110$, $E'_{2,0}=30$,
    $\epsilon'_{1}=\epsilon'_{2}=5$, $\mu'_{1}=\mu'_{2}=0.5$, $\Omega'_{2}=0.79$, and site-dependent parameters $\Omega_{1,n}=0.48/m_n$, $J'_{1,n}$, $J'_{2,n}$, $\chi'_{1,n}$ and $\chi'_{2,n}$ corresponding to $E_{0,1}=0.72$ eV, $E_{0,2}=0.2$ eV, $\epsilon_{1}=\epsilon_{2}=0.0329$ eV, $\mu_{1}=\mu_{2}=0.5$,  $\Omega_{2,n}=V^{2}\langle M\rangle/a^2$, $\Omega_{1,n}=V^{2}M_{n}/a^2$, $J_{1,n}$, $J_{2,n}$, $\chi_{1,n}$ and $\chi_{2,n}$
entering Equations (\ref{barrier}) and (\ref{barrier2}); $\sigma_{1,0}=\sigma_{2,0}=0.1$. a) DNA containing the specific CTTAAG recognition sites, b) randomized restriction sites AGCTTA, c) exchanging only one nucleotide with its complementary CATAAG, d) exchanging two nucleotides with their complementaries GTTAAC.}
 \label{37823725}
 \end{figure}

\begin{figure}[h!]
\includegraphics[width=0.45\columnwidth]{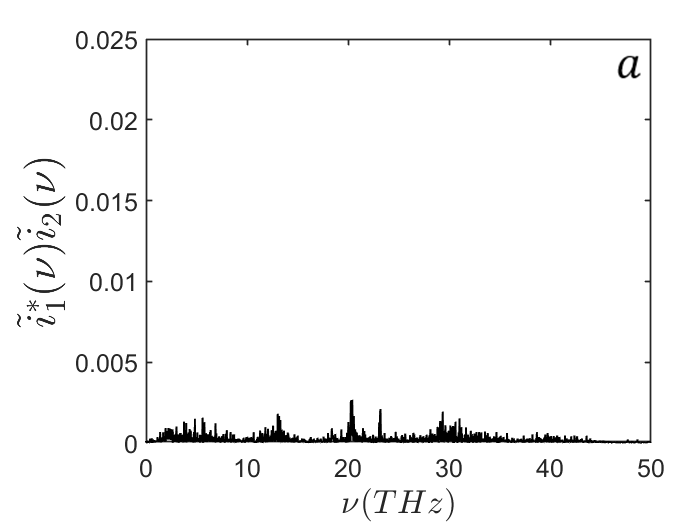}   
\includegraphics[width=0.45\columnwidth]{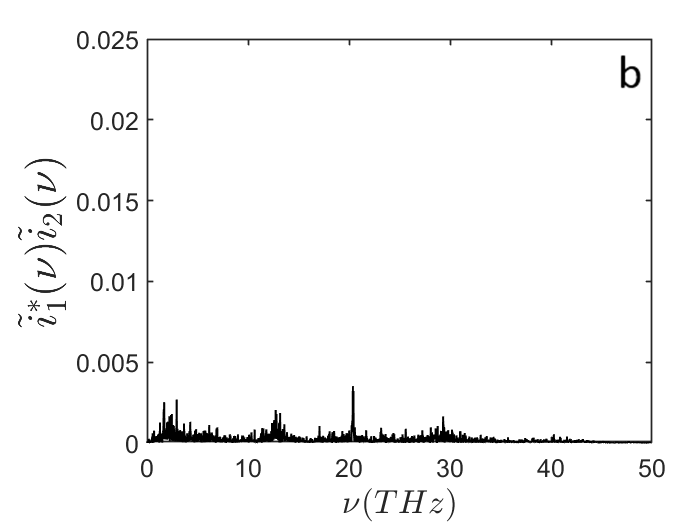}  
\includegraphics[width=0.45\columnwidth]{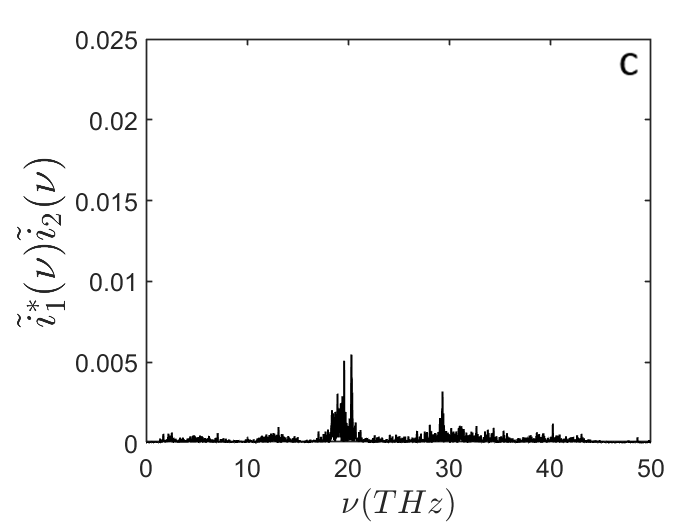} 
   \caption{ The cross frequency spectrum of the interaction between DNA and the EcoRI enzyme with same physical conditions as in Figure \ref{37823725}. a) The randomized restriction sites TCATGA, b)  exchanging only one nucleotide with its complementary CTTAAC, c) exchanging two nucleotides with their complementaries CATATG. The value of frequency units of $\nu$ is $10^{13}$s$^{-1}$.}
 \label{37823725bis}
 \end{figure}

\begin{figure}[h!]
 \includegraphics[width=0.45\columnwidth]{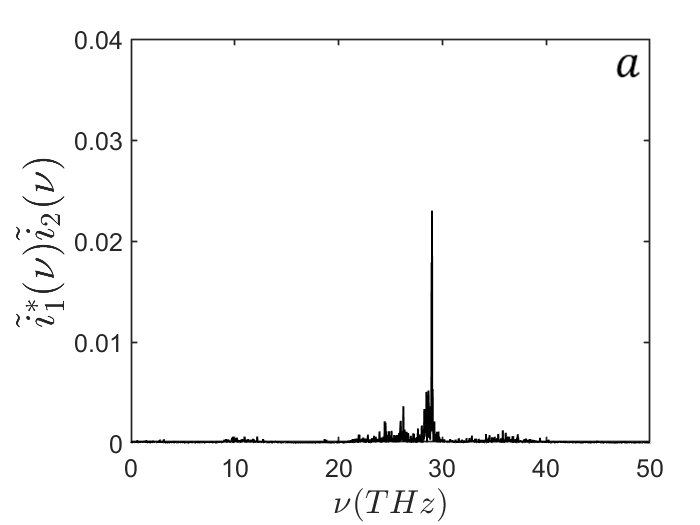}
 \includegraphics[width=0.45\columnwidth]{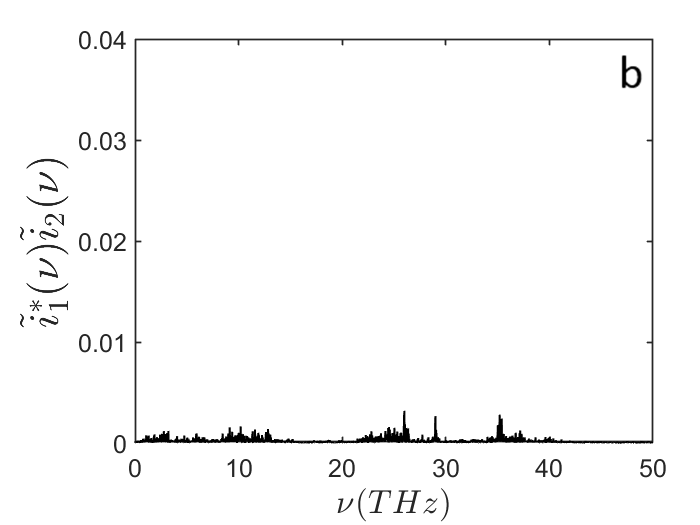}
 \includegraphics[width=0.45\columnwidth]{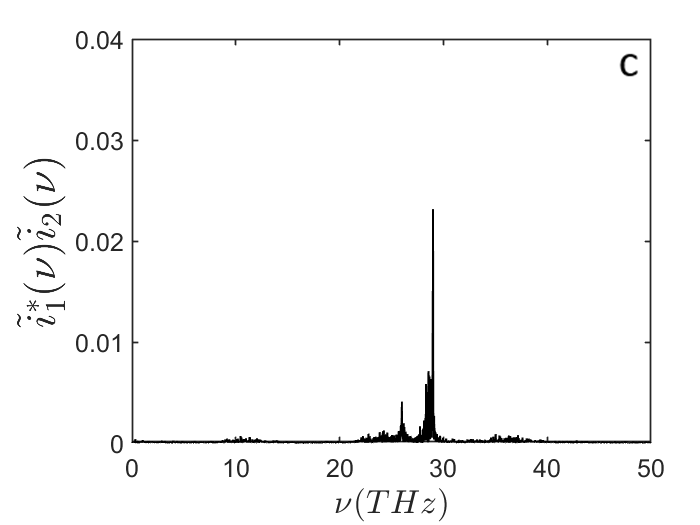} 
 \includegraphics[width=0.45\columnwidth]{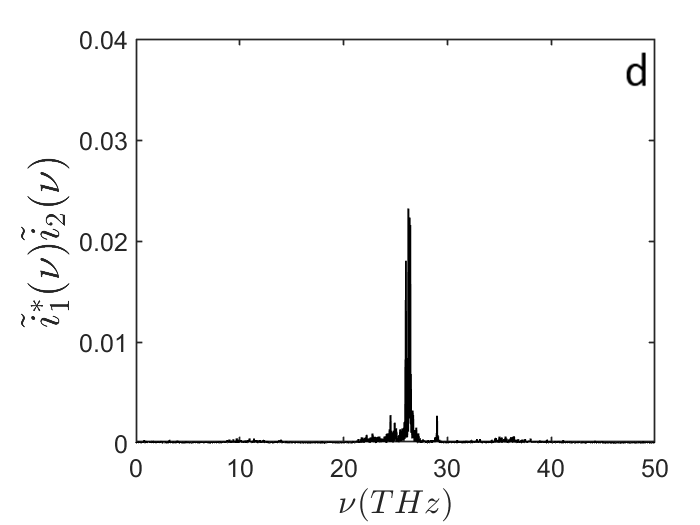}       
    \caption{The cross frequency spectrum of the interaction between DNA strand with $N_{1}=66$ nucleotides and the EcoRI enzyme with $N_{2}=276$ amino acids for the initial conditions:  $N_{0,1}=N/2$,  $N_{0,2}=N/3$, $E'_{1,0}=E'_{2,0}=129.17$,  $\Omega'_{1,n}=0.25$, $\Omega'_{2,n}=1.83/m_n$
    corresponding to  $E_{1,0}=E_{2,0}=0.85$ eV,  $\Omega_{1,n}=V^{2}M_{n}/a^2$ and $\Omega_{2}=18.3$ N/m. The other parameters are the same as in Figure (\ref{37823725}); a) DNA containing the specific recognition sites CTTAAG, b) randomized restriction sites TCATGA, c) exchanging only one nucleotide with its complementary site CTTATG, d) exchanging two nucleotides with their complementaries CTATAG.}
  \label{37163715}
  \end{figure}

\begin{figure}[htb]
  \includegraphics[width=0.45\columnwidth]{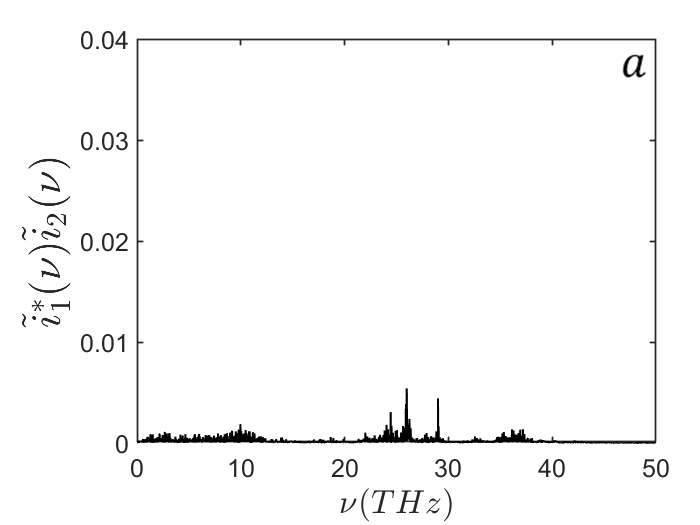}  
     \includegraphics[width=0.45\columnwidth]{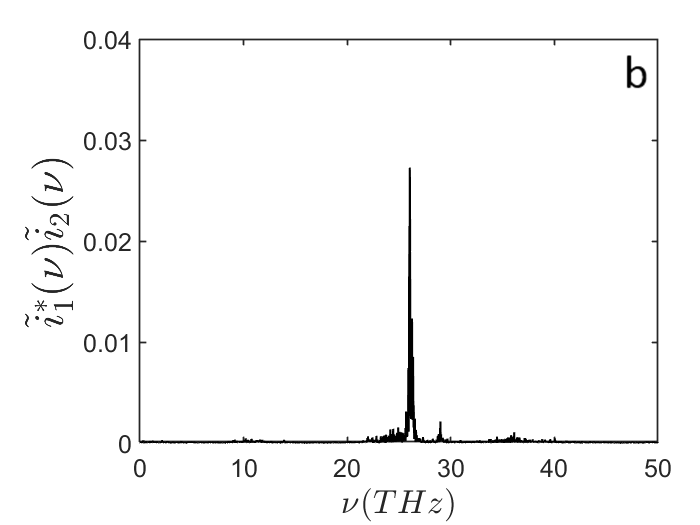}   
       \includegraphics[width=0.45\columnwidth]{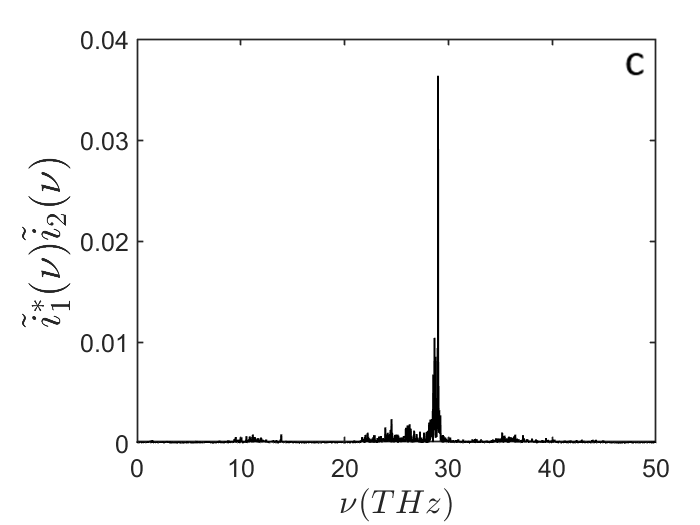}      
    \caption{ The cross frequency spectrum of the interaction between DNA and the EcoRI enzyme with same physical conditions as in Figure \ref{37163715}. a) randomized restriction sites AGATCT, b) exchanging only one nucleotide with its complementary site CATAAG, c) exchanging two nucleotides with their complementaries GTTATG. The value of frequency units for $\nu$ is $10^{13}$s$^{-1}$.}
  \label{37163715bis}
  \end{figure}

We have used an integration time step $\Delta t = 5\times 10^{-6}$ to work out our numerical simulations with a very good energy conservation and a typical relative error $\Delta E/E = 10^{-6}.$ The following analyses have reported the spectral properties of electron currents in the interaction of
a DNA fragment of $N=66$ nucleotides ( the $3'$-$5'$ direction of nucleotides shown in Figure. 1) and an EcoRI restriction enzyme of $N=276$ amino acids (displayed in Figure. 1) for different initial activation energies of electron $E_0$, various initial excitation sites $n_0$  of the electron in the probability amplitude (\ref{exc}), and  distinct forms of the phononic spring term $\Omega_{n}$. We study the Fourier spectrum of the electron current activated on a segment of DNA and also DNA-interacting enzyme, and, from now on, we use the index $1$ and $2$ for all the terms relative to DNA and EcoRI, respectively. Resorting to the standard probability current $j(x,t)$ of the electron wave function (\ref{Psi}), the electron density current is given by
\[
{j}({x},t) = \frac{e\hbar}{2m_e i}\left( \psi^\star\nabla\psi - \psi\nabla\psi^\star\right)\ ,
\]
hence the average electron current, in a spatially discretized form for numerical computation, is
\begin{eqnarray}
i_{1,2}(t)&=& \frac{1}{l_{1,2}} \int_{0}^{l_{1,2}} j_{1,2}(x,t) dx=\frac{e \hbar}{2N_{1,2}a_{1,2}m_{e}i}\nonumber\\
 &\times&\sum_{j=1}^{N_{1,2}}\Big(\Psi_{1,2}^{*}(x_j,t)\frac{\Psi_{1,2}(x_{j+1},t)-\Psi_{1,2}(x_{j-1},t)}{2}\nonumber\\ &-&\Psi_{1,2}(x_j,t)\frac{\Psi_{1,2}^*(x_{j+1},t)-\Psi_{1,2}^*(x_{j-1},t)}{2}\Big),
\end{eqnarray}
 where $l_{1,2}$ are the lengths, and $i_{1,2}$ are the currents flowing along the DNA fragment and the EcoRI enzyme macromolecules, respectively. In Figures. (\ref{37823725}),(\ref{37823725bis}),(\ref{37163715}),(\ref{37163715bis}), we have plotted the cross Fourier spectrum of the two currents $\tilde{i}^{*}_{1}(\nu)\tilde{i}_{2}(\nu)$ and studied whether the cleaving sequence CTTAAG of DNA, recognised by the EcoRI enzyme, entails some peculiarity associated to this kind of DNA-protein interaction. Fig. (\ref{37823725}) shows the behavior of the system when the excited electron on the DNA  has the initial energy $E_{1,0}=0.72$ eV and its wavefunction is initially centered at the site $n_{1,0}=N/2$, while for the restriction enzyme the initial excitation energy of the electron is $E_{2,0}=0.2$ eV localized at $n_{2,0}=N/3$. Besides, as we already discussed in Section \ref{numericalcomputations}, we consider the dimensionless expressions of the site-dependent phononic spring $\Omega'_{1,n}=0.48/m_n$ for the nucleotides and the constant term  $\Omega'_{2}=0.79$ for the amino acids. In panel (\ref{37823725})$a$ we see the very interesting phenomenon of a clear co-resonance around $20$ THz when the specific CTTAAG restriction sequence is taken into account. This result is in qualitative agreement, and  also in very good quantitative agreement, with the peak found by applying the RRM \cite{nota}. Another significant finding shown in panel (\ref{37823725})b is that the cross spectrum becomes completely spread when the recognition sites are randomly chosen as AGCTTA. Moreover, in panel (\ref{37823725})c, when we exchange just one nucleotide of the restriction sequence with its own complementary as CATAAG, the co-resonance undergoes a little alteration and broadens more by changing two nucleotides of the recognition sites in the form of GTTAAC presented in panel (\ref{37823725})d. In Figure  (\ref{37823725bis}) we assume the same initial and physical condition of Figure (\ref{37823725}) and evaluate the cross frequency spectrum with other substitution of nucleotides of the restriction sites. Again the sharp co-resonance peak in panel (\ref{37823725})$a$, is found to disappear in panel (\ref{37823725bis})a as a consequence of the randomization of the recognition sequence to TCATGA. Besides, the loss of the co-resonance displayed by the spectrum (\ref{37823725})$a$ is also found by changing only one nucleotide as CTTAAC in panel (\ref{37823725bis})b and two nucleotides as CATATG in panel (\ref{37823725bis})c.\\
  In Figure (\ref{37163715}) the results are obtained for different initial conditions, which confirms the robustness of the phenomenology previously seen in Figures (\ref{37823725}) and (\ref{37823725bis}). Here we assume the initial electronic activation energy $E_{1,0}=0.85$ eV given to  site $n_{1,0}=N/2$ in DNA macromolecule, and initial electronic activation energy in EcoRI enzyme  $E_{2,0}=0.85$ eV located in  $n_{2,0}=N/3$. Also, the dimensionless parameter of phononic spring in DNA fragment is assumed constant, with value $\Omega'_{1}=0.25$, and  in EcoRI enzyme it is considered site-dependent, with value  $\Omega'_{2,n}=1.83/m_n$. The sharp peak of co-resonant spectrum of the DNA-EcoRI interaction with the characteristic site restriction sites CTTAAG depicted in panel (\ref{37163715})$a$ happens around $29$ THz that broads entirely by choosing the randomized recognition sites TCATGA in panel (\ref{37163715})b. It is clear in panel (\ref{37163715})c that the well-peaked frequency spectrum ramifies very little by exchanging only one nucleotide of the recognized sites with its complementary as CTTATG and destroys somehow more when two nucleotides are exchanged as CTATAG seen in panel (\ref{37163715})d. Figure. (\ref{37163715bis}) shows the same initial condition of Figure. (\ref{37163715}) but with the different arrangement of nucleotides of the recognition sites. Taking the randomized sites AGATCT in the panel (\ref{37163715bis})e broads the co-resonance spectrum of the panel (\ref{37163715})$a$ while neither changing only one nucleotide considered as CATAAG in panel (\ref{37163715bis})f nor substituting two sites with their complementary ones as CATAAC in panel (\ref{37163715bis})g make the peak frequency spectrum broaden.

\section{Possible mechanisms mediating long-range DNA-protein interactions}
The results reported in the preceding section highlight the potential origin of selective electrodynamic interactions between DNA and proteins. In order to assess the actual relevance of the co-resonance of electron currents in biological contexts, a quantitative estimate of the strength of the implied interaction requires a similar strategy to the one reported analytically in \cite{preto}, as well as additional experimental data on the intensity of the currents and the possible mechanisms of their activation in a biological environment. These points will be tackled in future investigations, but in what follows we sketch possible scenarios that support electrodynamic mediation of DNA-protein interactions.

First, given two electron currents ${\bf j}^{(1)}(\mathbf{x},t)$ and ${\bf j}^{(2)}(\mathbf{x},t)$ representing those of DNA and \textit{Eco}RI, respectively,
\[
{\bf j}^{(1,2)}(\mathbf{x},t) = \frac{e\hbar}{2m_e i}\left( \psi^\star\vec{\nabla}\psi - \psi\vec{\nabla}\psi^\star\right)\ ,
\]
and according to the D'Alembert equations (in Gaussian units and Lorenz gauge),
\[
\square^2{\bf A}^{(1)}(\mathbf{x},t) =(4\pi/c) {\bf j}^{(1)}(\mathbf{x},t)\ ,
\] 
and
\[
\square^2{\bf A}^{(2)}(\mathbf{x},t) =(4\pi/c) {\bf j}^{(2)}(\mathbf{x},t),
\] 
the mutual interaction is described by the coupling terms 
\[
{\bf j}^{(2)}(\mathbf{x},t)\cdot{\bf A}^{(1)}(\mathbf{x},t) \; \; \; \; \; \; \; \; \; \;{\rm and}\; \; \; \; \;\; \; \; \; \;
{\bf j}^{(1)}(\mathbf{x},t)\cdot{\bf A}^{(2)}(\mathbf{x},t) \ .
\]
Since the D'Alembert equation is linear, the vector potential inherits the spectral properties of the current that generates it. As a  consequence, the 
co-resonance between the two currents ${\bf j}^{(1)}(\mathbf{x},t)$ and ${\bf j}^{(2)}(\mathbf{x},t)$ entails the largest values of the time averages of the interaction 
energies. 
\medskip

Second, intriguing connections exist between the models presented above, which describe electronic motions along a given DNA sequence and a given protein sequence, and the coordinated electronic fluctuations that arise from van der Waals many-body dispersion forces \cite{kurian2016, tkatchenko2017, kurian2018, gori2022} in a variety of molecular contexts. Specifically, productive insights have emerged from attempts to unify atomistic, continuum, and mean-field treatments in the quantum electronic behaviors of DNA and proteins in water \cite{kurian2018, gori2022, kurianchiral, tkatchenko2019, kurian2022}. Even in the presence of thermally turbulent aqueous environments, it has been shown that these collective electronic dispersion correlations can persist at several nanometers from the protein-water interface, and these correlations are energetically relevant for protein-folding processes at the microsecond scale \cite{tkatchenko2019}, and likely for even longer times \textit{in vivo}. 
\begin{figure}[htb]
\centering
\includegraphics[width=0.5\textwidth]{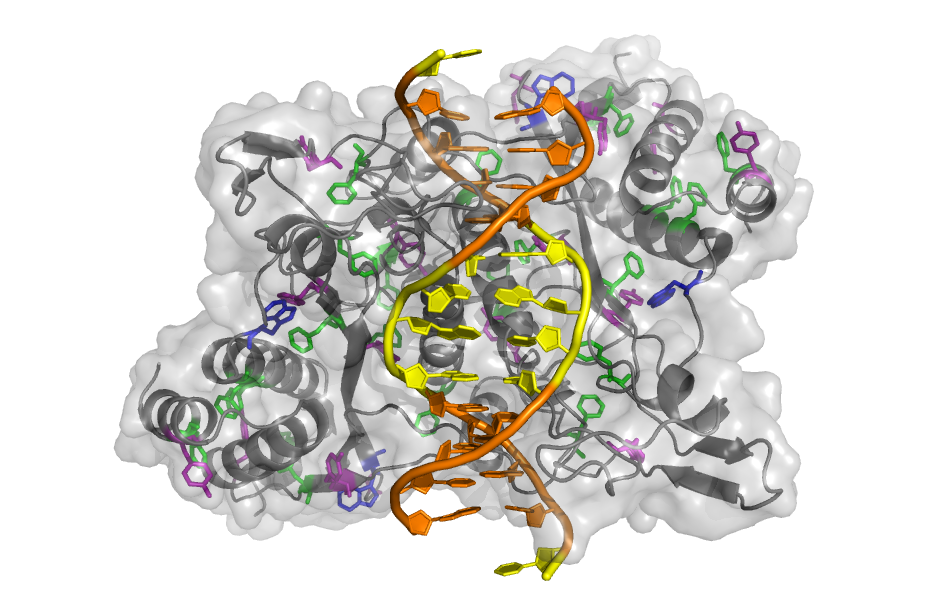}
\includegraphics[width=0.5\textwidth]{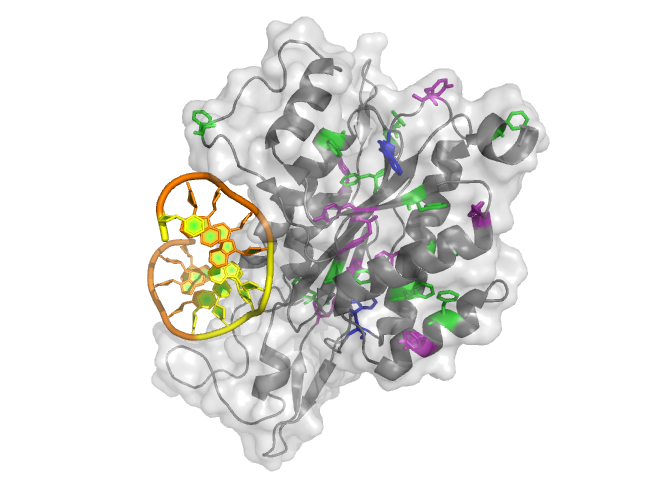}
\caption{Aromatic network in \textit{Eco}RI-DNA complex. Tryptophan (blue), tyrosine (purple), and phenylalanine (green) form correlated electronic dispersion networks in \textit{Eco}RI, shown here in the top panel bound to its double-stranded DNA substrate, with adenine-thymine (yellow) and cytosine-guanine (orange) base pairs highlighted. Other amino acids (gray) are displayed in the context of their secondary structures within the enzyme, and in the bottom panel only one of the two \textit{Eco}RI dimers is shown for clarity, to showcase the $\pi-\pi$ stacking of the DNA bases. Image of \textit{Eco}RI (PDB ID: 1CKQ) at 1.85 \r{A} resolution created with PyMOL and adapted from \cite{kurian2018}.}
\label{fig:aroEco}
\end{figure}

Kurian and coworkers \cite{kurian2016, kurian2018} have additionally shown that such collective electronic (quantum harmonic oscillator) modes are suitably fine-tuned for the synchronized catalysis of two phosphodiester bonds ($\sim0.46$ eV), and that the palindromic mirror symmetry of the double-stranded DNA target sequence recognized by the enzyme (see Figure \ref{fig:aroEco}) allows for conservation of parity in the symmetric, site-specific cleavage of both DNA strands. By considering the radiative field $\mathbf{E}$ created by the collective electronic fluctuation modes in the DNA target sequence, a nonvanishing polarization density emerges spontaneously in the orientational correlations of the water dipole network through the interaction Hamiltonian $H = -\mathbf{d}_e \cdot \mathbf{E}$, where $\mathbf{d}_e$ is the permanent electric dipole moment for a single water molecule.

Following standard treatments in quantum optics \cite{gerryknight}, this interaction between the DNA radiative field and the surrounding (quasi-continuous) water dipole field can be written in the form of a Jaynes-Cummings-like Hamiltonian that scales with the number of water molecules $N$ as 
\begin{equation}
H_{\text{int}} = \hbar \sqrt{N} \gamma (a^{\dagger}S^- + aS^+),
\label{JCHam}
\end{equation}
where $\gamma$ is the coupling constant proportional to the matrix element of the molecular dipole moment and inversely proportional to the volume square root, $a^{\dagger}$ and $a$ are the creation and annihilation operators, respectively, for the DNA radiative electric field $\mathbf{E}$, and $S^+$ and $S^-$ are the raising and lowering operators, respectively, for the collective water dipole state. The quasi-continuum water dipole ``field'' thus takes the place of the $N$ two-level systems described in the Tavis-Cummings model \cite{TC}.
 
It should be noted that the coupling $\sqrt{N} \gamma$ in Equation (\ref{JCHam}) between the DNA radiative field and the collective water orientational state levels \cite{kurian2018} scales with the square root of the water density $\rho$, which varies with temperature and pressure. However, if we consider that the number of water molecules in a (cubic) domain encompassed by infrared wavelengths $\gtrsim1\,\mu$m exceeds 10 billion, such sufficiently large $N$ for the collective state can provide a protective gap against thermalization ($k_B T \approx 0.02$ eV at physiological temperatures) for the long-range correlations we consider. Furthermore, the spontaneous breakdown of phase symmetry generates a field polarization (in the so-called ``limit cycle'' regime) that preserves gauge invariance by dynamical coherence between the matter quasi-continuum field (DNA, water, enzyme) and the phase-locked electromagnetic field (radiative field from DNA, water, enzyme).

As a toy model, we use Faraday's law of induction for the DNA double helix, considered here as a long solenoid with radius $R$, $n$ turns per unit length, and current along the backbone varying as $I = I_0 e^{-\alpha t}$, where $\alpha$ is in general complex. For distances from the longitudinal axis $r > R$ outside the helix-solenoid, we can estimate the induced electric field $\mathbf{E}(r,t)$ tangent to a circular path surrounding the cylindrically symmetric system:
\begin{equation}
|\mathbf{E}(r,t)| = \frac{\Omega}{2} \frac{|e^{-\alpha t}|}{r},
\label{helixE}
\end{equation}
where $\Omega = |\alpha| \mu n I_0 R^2$ and $\mu$ is the magnetic permeability in water. From Equation (\ref{helixE}) we can thus derive the creation and annihilation operators $a^\dagger, a$ for the radiative field in the interaction Hamiltonian of Equation (\ref{JCHam}).

The resulting interaction energies range between $\sim0.1-1$ eV, populating bands in the infrared spectrum between $0 < \nu < 1000 \text{ cm}^{-1}$, which overlaps with the energy scale of the collective electronic fluctuation modes in the DNA target sequence and in the enzyme when taken separately, but remains distinct from the more energetic intramolecular vibrations and purely electronic transitions of individual water molecules. These collective electronic fluctuation modes in the $0.1-1$ eV range do not couple to the rotational quantum transitions of individual water dipoles (meV scale), but rather to the emergent polarization modes present in the collective dipole network. The spectroscopic peaks for liquid water also lie completely within this range.

Chiral sum frequency generation spectroscopy experiments \cite{chiralspine} have demonstrated the existence of a chiral water superstructure surrounding DNA under ambient conditions, thereby confirming that the chiral structure of DNA can be imprinted electrodynamically on the surrounding solvent. These experiments have also shown that some sequence-specific fine structure persists in this chiral spine of hydration, providing a mediating context for DNA target sequence recognition by various proteins.

\section{Concluding remarks}   
The aim of the present paper is twofold. First, inspired by the Resonant Recognition Model (RRM), we wanted to tackle biomolecular resonant interactions by resorting to a widely used electron-phonon Hamiltonian applied to alternating currents along the backbone of specific DNA target sequences in the second quantization framework. Second, the work here reported contributes to the still open discussion of long-distance electrodynamic intermolecular interactions, which have recently been demonstrated experimentally \cite{SciAdv,buchanan}.

Regarding the first aim,  the above mentioned model was applied to the pair of partners of the biochemical reaction involving a DNA fragment and a restriction enzyme, \textit{Eco}RI, that binds to a specific target subsequence of the DNA fragment to cleave it. 
The interaction energies of an electron with the sequence of nucleotides composing a specific DNA fragment on the one side, and the interaction energies of an electron with the sequence of amino acids composing the \textit{Eco}RI enzyme on the other side, yield two numerical sequences. The product of their Fourier spectra, or cross-spectrum, displays a sharp peak.
The peak so found qualitatively witnesses to the specific relationship between the two biomolecules, though the physics behind this co-resonance still needs to be clarified. Such a clarification is provided by the co-resonance of the time-domain Fourier spectra of the alternating electron currents moving along the DNA and enzyme, respectively. These currents are worked out through second quantization dynamical models describing the electron-phonon coupling, which are derived from standard Davydov and Holstein-Fr\"{o}hlich treatments \cite{standard,froehpolaron,holstein}. The remarkable finding is the disappearance of the co-resonance peak when the six-base-pair (bp) target recognition subsequence GAATTC on the DNA is randomized in different ways.  
Regarding the second aim of the paper, the prospective relevance for biology of long-range selective and attractive intermolecular interactions was discussed in the Introduction and has recently been given experimental confirmation \cite{SciAdv} in the presence of collective intramolecular oscillations. The question naturally arises whether the electronic degrees of freedom of electrodynamically interacting molecules can offer alternative or complementary mechanisms to activate such long-range intermolecular forces. We have presented a first step in this second direction, and the remarkable finding mentioned above motivates further investigations. In fact, at present we have considered the motion of a single electron, but we can think that under suitable excitation processes (for example, under repeated ATP hydrolysis events or near an ionic channel) definitely stronger currents can be activated, producing either direct electrodynamic current-to-current interactions, or, as intriguingly proposed in \cite{kurian2018} and discussed in the preceding section, water-mediated electrodynamical interactions between the radiative field emerging from electronic fluctuational motions in DNA and in protein, and the water dipole (matter) field in the quasi-continuum limit.
Finally, the observed sequence-dependent co-resonance phenomenology for the chosen biochemical model is suggestive of a potentially rich variety of selective electrodynamic interactions of a more general kind, including, for example, those between DNA molecules and transcription factors undergoing electron-phonon excitation. 

 \section{Data availability}
The Matlab scripts used to produce the data sets of the current study are available at  \href{https://doi.org/10.5281/zenodo.10593456}{10.5281/zenodo.10593456} .
 
\section*{Acknowledgments}
E.Faraji warmly thanks the Fondazione Cassa di Risparmio di Firenze for having co-funded her PhD fellowship.
P.Kurian acknowledges support from the U.S.-Italy Fulbright Commission, National Science Foundation, Whole Genome Science Foundation, and Alfred P. Sloan Foundation.
M.Pettini participated in this work within the framework of the project MOLINT which has received funding from the Excellence Initiative 
of Aix-Marseille University - A*Midex,  a French "Investissements d'Avenir" programme. 
R. Franzosi acknowledges the support from the Research Support Plan 2022 – Call for applications for funding allocation to research projects curiosity driven (F CUR) – Project “Entanglement Protection of Qubits’ Dynamics in a Cavity” – EPQDC, the support by the Italian National Group of Mathematical Physics (GNFM-INdAM) and the financial support by INFN Pisa for this activity. S.Mancini acknowledges financial support from "PNRR MUR project PE0000023-NQSTI". E. Floriani, M.Pettini and V.Calandrini acknowledge support by the European Union’s Horizon Research and Innovation Programme under Grant Agreement No. 964203 (FET-Open LINkS project).
Inspiring exchange of scientific information with Irena  and Drasko Cosic are warmly acknowledged.

\appendix
\section{}\label{mut}
\begin{figure}[h!]
  \includegraphics[width=0.45\columnwidth]{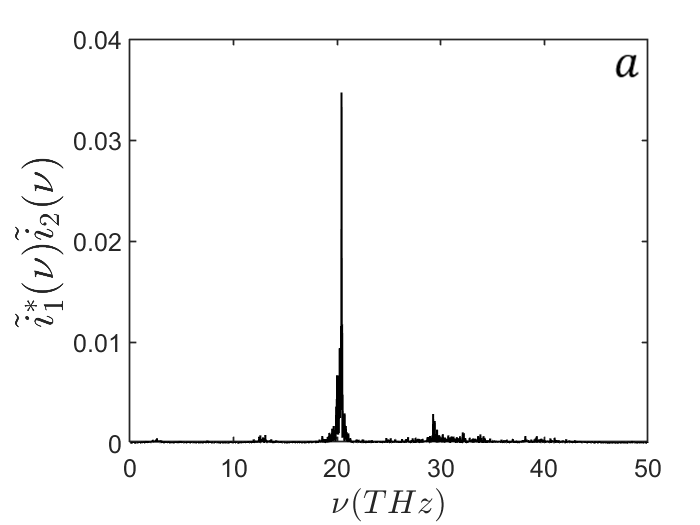}  
     \includegraphics[width=0.45\columnwidth]{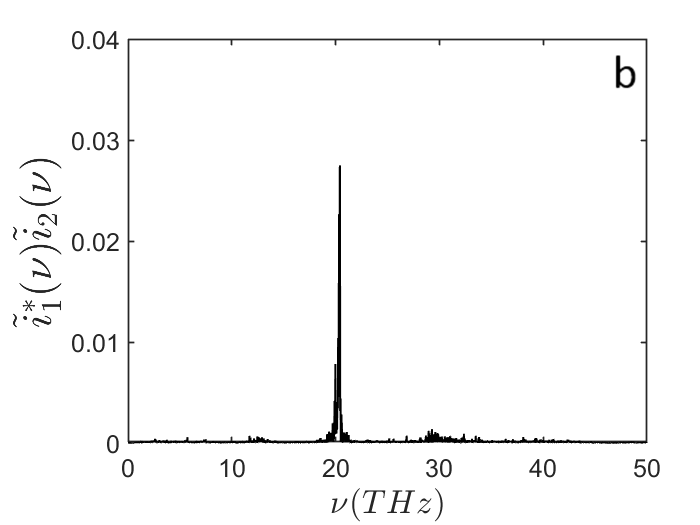}   
       \includegraphics[width=0.45\columnwidth]{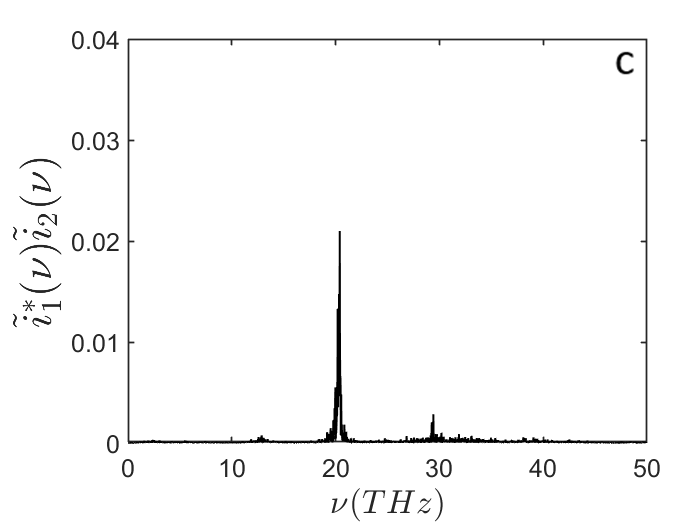}   
       \includegraphics[width=0.45\columnwidth]{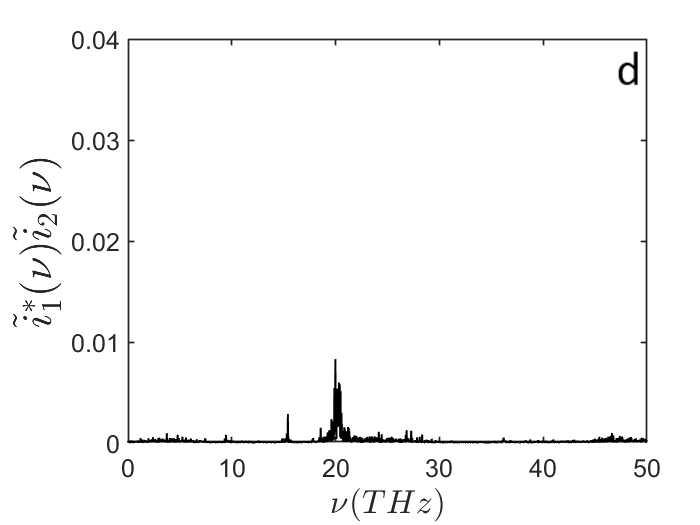}
\caption{ {Cross frequency spectra between the original substrate and the \textit{Eco}RI mutants with the single point mutations;  a) $Ala^{138}\to Thr$; b) $Glu^{192}\to Lys$; c) $His^{114}\to Tyr$; and d) $Asp^{91}\to Asn$. The initial conditions are the same of Figure~\ref{37823725}.}}
  \label{fig:mutations}
  \end{figure}

 {To further asses the coherence of the model, we evaluate the cross frequency spectrum between the substrate and \textit{Eco}RI, upon single point mutations on the primary sequence of the protein. Using the same initial conditions of Fig.~\ref{37823725}, we specifically test the mutations $Ala^{138}\to Thr$, $Glu^{192}\to Lys$, $His^{114}\to Tyr $, and $Asp^{91}\to Asn$, where $x^{y}\to z$ indicates that the amino acid $x$ at position $y$ in the primary sequence of the protein has been replaced by the amino acid $z$. The first three mutations, belonging to the group of the so-called promiscuous mutations, have been proved to produce mutants able to bind both the cognate sequence CTTAAG  and miscognate sites (differing by a single base-pair from the canonical one)~\cite{sapienza2005}, whilst the last mutation has been shown to essentially reduce the catalytic activity of \textit{Eco}RI ~\cite{grabowski1995}. Panels a,b,c of Fig.~\ref{fig:mutations} show that the promiscuous mutations still yield a sharp co-resonant peak. On the contrary, the mutation $Asp^{91}\to Asn$ produces a sizeable decrease of the peak (panel d of Fig.~\ref{fig:mutations}). These findings are coherent with the experimental data.}

\end{document}